\documentclass[aps,prc,superscriptaddress,twocolumn,nofootinbib]{revtex4}
\usepackage{amsmath}
\usepackage{graphicx}
\usepackage[T1]{fontenc}
\usepackage{epsfig}
\usepackage{amssymb}
\usepackage{mathrsfs}
\usepackage{mathcomp}
\usepackage{amsmath}

\newcommand{\beq}{\begin{eqnarray}}
\newcommand{\eeq}{\end{eqnarray}}

\begin{document}
\title{Eccentricity and elliptic flow in proton--proton collisions from
  parton evolution}
\author{Emil Avsar}
\affiliation{
104 Davey Lab, Penn State University, University Park, 16802 PA, USA}
\author{Christoffer Flensburg}
\affiliation{
Dept. of Theoretical Physics, S\"olvegatan 14 A, S223 62 Lund, Sweden}
\author{Yoshitaka Hatta}
\affiliation{
Graduate School of Pure and Applied Sciences, University of Tsukuba,
Tsukuba, Ibaraki 305-8571, Japan}
\author{Jean-Yves Ollitrault}
\affiliation{
CNRS, URA2306, IPhT, Institut de physique th\'eorique, CEA Saclay, F-91191
Gif-sur-Yvette, France}
\author{Takahiro Ueda}
\altaffiliation[Present address: ]{
Institut f\"ur Theoretische Teilchenphysik,
Karlsruhe Institute of Technology (KIT),
D-76128 Karlsruhe, Germany}
\affiliation{
Graduate School of Pure and Applied Sciences, University of Tsukuba,
Tsukuba, Ibaraki 305-8571, Japan}

\date{\today}
\begin{abstract}
It has been argued that high--multiplicity proton--proton collisions at
the LHC may exhibit collective phenomena usually studied in the
context of heavy--ion collisions, such as elliptic flow. We study this
issue using DIPSY---a
Monte Carlo event generator based on the QCD dipole model.
We calculate the eccentricity of the transverse area
defined by the spatial distribution of produced gluons.
The resulting elliptic flow is estimated to be about $6\%$,
comparable to the value in nucleus--nucleus collisions at RHIC and the LHC.
Experimentally, elliptic flow is inferred from the azimuthal
correlation between hadrons, which receives contributions from
collective flow, and from various other effects referred to as
``nonflow''. We discuss how to identify in experiments the signal of flow in the presence of large
nonflow effects.
\end{abstract}
\maketitle

\section{Introduction}
\label{s:intro}

Elliptic flow is one of the most important
phenomena observed in ultrarelativistic nucleus--nucleus
collisions~\cite{Ackermann:2000tr,Adler:2003kt,Alver:2006wh}.
An Au-Au collision at RHIC produces several thousands of
particles.
If interactions among these particles are strong enough
they expand collectively like a fluid, and elliptic flow is a probe of
this collective behavior~\cite{Ollitrault:1992bk}.
The fluid picture is a macroscopic one, which is generally valid for a
large system.
For a system as small as a
nucleus, it is an idealization which must be amended in order to
quantitatively understand experimental data~\cite{Drescher:2007cd}.
The system formed in proton--proton ($pp$) collisions is even smaller. Yet the
possibility has been raised that elliptic flow may be seen in
 $pp$ collisions at the
LHC~\cite{Luzum:2009sb,Bautista:2009my,Chajecki:2009es,Prasad:2009bx,d'Enterria:2010hd,Bozek:2009dt,
Ortona:2009yc,
CasalderreySolana:2009uk,Pierog:2010wa}. This is actually a quite nontrivial problem which can only be addressed with a proper understanding of the proton wavefunction at high energy from QCD, whereas most of the preceding works~\cite{Luzum:2009sb,Prasad:2009bx,d'Enterria:2010hd,Bozek:2009dt,Ortona:2009yc,
CasalderreySolana:2009uk}  are based on rather primitive models of the proton.  [See, however, \cite{Bautista:2009my,Pierog:2010wa}.]
In this letter, we study this issue using a full Monte Carlo (MC) model of the
collision which implements the BFKL--type evolution of  structure
functions, multiple collisions, the partonic shower and the subsequent hadronization. This
model is briefly described in Sec.~\ref{s:model}.

An obvious obstacle to develop collective phenomena in $pp$ collisions is the
 low multiplicity of hadrons in the final state. This may be
overcome by triggering on high--multiplicity events. Indeed, it has already been
observed in the 7 TeV run at the LHC \cite{Aamodt:2010pp,Collaboration:2010gv} that
the multiplicity distribution has a broad tail reaching out to $\frac{dN_{ch}}{d\eta}>30$,
and this will be further pronounced  in future runs at 14 TeV. Such high--multiplicity events originate from
  upward fluctuations in the gluon multiplicity inside the proton and the subsequent multiple
 gluon--gluon scatterings. The fluctuations in the distribution of gluons in the transverse plane
 then generates nonzero eccentricity of the interaction region even in collisions at vanishing impact parameter.\footnote{Previous works considered the fluctuation of `hot spots' \cite{CasalderreySolana:2009uk} and `flux tubes' \cite{Pierog:2010wa}. There the  transverse distribution of these objects was assumed to be random. In our case the transverse distribution of gluons is  governed by the QCD evolution. }
Assuming hydrodynamic evolution for these high--multiplicity events,
 we estimate the magnitude of the resulting elliptic flow in Sec.~\ref{s:epsilon}.


Experimentally, elliptic flow is not measured directly, but inferred from azimuthal
correlations between the produced particles. These correlations are
partly due to elliptic flow, partly due to other effects referred to
as ``nonflow''~\cite{Dinh:1999mn}.
Nonflow correlations are sizable for peripheral nucleus--nucleus collisions at
RHIC~\cite{Ollitrault:2009ie}, and one expects them to be even larger
in $pp$ collisions, making it a challenging task to disentangle the flow contribution.
In Section IV, we stress the necessity to look at higher--order cumulants of azimuthal correlations, and suggest how to identify flow in light of the experimental and MC results.

\section{The model}
\label{s:model}

Our calculations  are based on the MC implementation of
the dipole model developed in Lund \cite{Avsar:2005iz, Avsar:2006jy, Avsar:2007xg,
Flensburg:2008ag,Flensburg:2011kk,Flensburg:2010jk}.
The  dipole model by Mueller  \cite{Mueller:1993rr,Mueller:1994gb}
realizes the leading--order BFKL evolution of gluons in the transverse
coordinate space, which is ideally
suited for the computation of the eccentricity. It is known that the BFKL evolution generates
large event--by--event fluctuations in the gluon multiplicity \cite{Salam:1995zd} as well as
characteristic
spatial distributions and correlations in the transverse plane \cite{Hatta:2007fg,Avsar:2008ph}.
Both of these effects are important in properly estimating the eccentricity.



For phenomenology, the original leading--order formulation is impractical, and over the years there have been many improvements of the model.  These include the running of the coupling, energy--momentum conservation, saturation effects, and confinement effects at large dipole separations. [For details, see
\cite{Avsar:2005iz, Avsar:2006jy, Avsar:2007xg, Flensburg:2008ag}.]   The parameters of the model are determined so
that they reproduce a few observables (such as the
total $pp$ cross section) at some energy. Predictions can then be made for various other observables at different
energies without any further tuning of the parameters.

The MC code that we actually use in the following, called DIPSY \cite{Flensburg:2011kk,Flensburg:2010jk}, is  the most advanced version by the Lund dipole team which
has access to all the exclusive final states. In this framework, a typical  high--multiplicity
event looks as follows: Before the scattering each proton develops a cascade  of gluons (or equivalently, dipoles) spread
in rapidity and the transverse plane. These gluons, mostly soft ones, then undergo multiple scatterings. The
evaluation of the non--diffractive scattering amplitude for the two cascades reduces to that for individual pairs of dipoles, allowing
DIPSY to decide on an event--by--event basis which dipoles interact. It is then possible to trace the
interacting parton chains back from the interactions, and the initial state radiation can be identified.
 All emissions not connected to the interacting chains are reabsorbed as virtual fluctuations.
 The initial state radiation is then passed to \protect{ARIADNE}
\cite{Lonnblad:1992tz}
that further splits the dipoles with timelike emissions. After that the dipoles hadronize through the string
fragmentation model in \protect{PYTHIA} \cite{Sjostrand:2006za,Andersson:1983ia}, giving
the observable final states.

\section{Eccentricity and the elliptic flow}
\label{s:epsilon}



In a nucleus--nucleus collision, the participant eccentricity $\epsilon_{\rm part}$
is defined from the positions $(x,y)$ of
participant nucleons within the nucleus~\cite{Alver:2006wh}:
\beq
\epsilon_{\rm part} \equiv \frac{\sqrt{(\sigma_y^2-\sigma_x^2)^2+4\sigma^2_{xy}}}
{\sigma_y^2+\sigma_x^2},
\label{epspart}
\eeq
where
\beq
\sigma_x^2 &=& \{x^2\} - \{x\}^2,  \nonumber \\
\sigma_y^2 &=& \{y^2\} - \{y\}^2,  \nonumber \\
\sigma_{xy} &=& \{x \, y\} - \{x\} \{y\},
\label{sigmadefs}
\eeq
and the brackets $\{\cdots \}$ denote averaging over
the participants
in a given event.  We shall be interested in the quantities $\epsilon\{2\}$ and $\epsilon\{4\}$ defined by
\beq
\epsilon \{2\}\! &\equiv& \!\sqrt{\langle \epsilon_{\rm part}^2 \rangle }\,,
\label{eps2} \\
\epsilon\{4\} \! &\equiv&\! ( 2 \langle \epsilon_{\rm part}^2 \rangle^2 - \langle \epsilon_{\rm part}^4 \rangle
)^{1/4}\,,
\label{eps4}
\eeq
 where $\langle \cdots\rangle$ denotes averaging over events in a given centrality bin.
  Hydrodynamic evolution linearly relates $\epsilon\{n\}$ and the corresponding elliptic flow $v_2\{n\}$
  measured from the $n$--particle azimuthal correlation \cite{Bhalerao:2006tp}.
  [See, however, a recent study \cite{Qiu:2011iv} which  suggests a possible mixing
of different harmonics due to fluctuations.]
  An empirical formula which works at RHIC is
\cite{Drescher:2007cd}
\beq
v_2\{2\} = \epsilon\{2\}  \left ( \frac{v_2}{\epsilon} \right )_{{\rm hydro}} \, \frac{1}{1+\frac{\lambda}{K_0}
\frac{\langle S \rangle }{ \left\langle \frac{dN}{d\eta}\right\rangle}}\,,
\label{hydroform}
\eeq
 and a similar relation between $v_2\{4\}$ and $\epsilon\{4\}$.
In (\ref{hydroform}), $(v_2/\epsilon)_{{\rm hydro}}\approx 0.2$ is the ideal hydrodynamics result
and the parameter $\lambda/K_0 = 5.8$ fm$^{-2}$ measures the degree of incomplete equilibration. $S$ is the area of the overlap region calculated as
\beq
S  = 4 \pi \sqrt{\sigma_x^2\sigma_y^2- \sigma_{xy}^2}\,,
\label{areaS}
\eeq
and $\frac{dN}{d\eta}\approx 1.5\frac{dN_{ch}}{d\eta}$ is the total hadron rapidity distribution. [We neglect the small difference between the rapidity and the pseudorapidity.]


  In $pp$ collisions, we employ the same formulae \eqref{epspart}--\eqref{eps4}, with the  averaging in \eqref{sigmadefs}  performed for the ``liberated'' gluons, i.e.,
those in the initial state radiation.
In doing so, we apply a
rapidity cut such that only gluons which are separated from the beam directions by more
than 2 units of rapidity are included in the averaging (\ref{sigmadefs}).\footnote{We could have included only gluons which are centrally produced, say, within $|\eta|<1$. However, we find it more reasonable to average over a wider range in rapidity in order to make up for the lack of the final state radiation in the eccentricity calculation. Noncentral gluons are connected to the central ones by color stings, and the early stage final state radiation from the noncentral gluons will contribute to the central energy density (and vice versa) before hydrodynamics starts to operate.  }   The validity of the use of (\ref{hydroform}) in $pp$ collisions at the LHC is \emph{a priori} not clear and requires an explanation. We first note that hydrodynamic simulations show that $v_2/\epsilon$ as a function of $\frac{1}{S}\frac{dN}{d\eta}$ falls essentially on the same curve both at RHIC and at the LHC  in spite of the difference in energy by a factor of about 14 \cite{Luzum:2009sb}, and both for Au--Au and Cu--Cu at RHIC although they differ in size by a factor of two \cite{Drescher:2007cd}. [The parameters $(v_2/\epsilon)_{\rm hydro}$ and $\lambda/K_0$ in principle depend on the temperature, but their changes are small due to the softness of the QCD equation of state.]  We then recall the general argument that the applicability of hydrodynamics is controlled by the dimensionless parameter $\alpha \equiv \frac{\lambda}{K_0}\frac{S}{dN/d\eta}$ rather than the system size. In our simulations,  the necessary condition $\alpha <1$ is well satisfied in a broad $N_{ch}$ range even  after allowing for some uncertainty in the parameter $\lambda/K_0$. On the other hand, it is hard to imagine hydrodynamic behaviors in systems smaller than $S\sim 1\, {\rm fm}^2$ which roughly sets the border between the hadronic and nuclear scales. We thus expect that (\ref{hydroform}) can be marginally applied for $S > 1\, {\rm fm}^2$, and this condition is better satisfied in high--multiplicity events (see below).

For the actual evaluation of $v_2$,  we propose the following slight improvement  of (\ref{hydroform})
\beq
(v_2\{2\})^2 =  \left ( \frac{v_2}{\epsilon} \right )^2_{{\rm hydro}} \,
\left\langle \frac{\epsilon^2_{\rm part} }{\bigl(1+\frac{\lambda}{K_0}
\frac{ S}{ dN/d\eta }\bigr)^2} \right\rangle \,,
\label{hydroform2}
\eeq
 and similarly,
 \beq
 (v_2\{4\})^4 = \left(\frac{v_2}{\epsilon}\right)^4_{{\rm hydro}} \Biggl\{ 2 \left\langle \frac{\epsilon^2_{\rm part}}{\bigl(1+\frac{\lambda}{K_0}\frac{S}{dN/d\eta} \bigr)^2}\right\rangle^2 \nonumber \\
 - \left\langle \frac{\epsilon^4_{\rm part}}{\bigl(1+\frac{\lambda}{K_0}\frac{S}{dN/d\eta} \bigr)^4}\right\rangle \Biggr\}\,. \label{hydroform3}
 \eeq
The reason is that in $pp$ collisions the eccentricity $\epsilon$ and the area $S$ fluctuate widely even at fixed $dN/d\eta$.
Equations (\ref{hydroform2}) and (\ref{hydroform3}) nicely captures this event--by--event correlation between $\epsilon$ and $S$.
Note that it is the squared value
$(v_2\{2\})^2$ (and also $(v_2\{4\})^4$) that directly comes out of the experimental measurement of
flow via multiparticle correlations~\cite{Ollitrault:2009ie}
\beq
v_2^2\{2\} \!\!\! &=&\!\!\! \bigl\langle \bigl\{\cos (2(\phi_i - \phi_j))\bigr\}
\bigr\rangle \,,  \label{v2def}\\
v_2^4\{4\} \!\!\! &=&\!\!\!
 2(v_2\{2\})^4 \!-\! \langle \{ \cos(2(\phi_i \!+\!\phi_j\!-\!\phi_k \!-\!
\phi_l))\}\rangle,  \label{v4def}
\eeq
where $\phi_i$ is the azimuthal angle of the $i$-th outgoing particle and averaging
 over all pairs (and 4-plets) satisfying some cut requirements is implied.
In the case of nucleus--nucleus collisions, the fluctuations are rather small so that (\ref{hydroform2}) essentially reduces to the previous formula (\ref{hydroform}).

We have generated $pp$ events at $\sqrt{s}=$7 TeV and 14 TeV with randomly chosen impact parameter $\vec{b}$, and classified events in bins of the charged particle multiplicity $N_{ch}$. The averaging $\langle \cdots \rangle$ has been taken in each bin.  Unlike in nucleus--nucleus  collisions, in $pp$ collisions the impact parameter is not measurable, and there is no simple scaling between the centrality and the multiplicity (not even between the centrality and the effective area $S$) because of the fluctuations. Still, the majority of high--multiplicity events in our MC simulations comes from
collisions with $b\approx 0$.

\begin{figure}[ht]
\resizebox{\linewidth}{!}{\rotatebox{-90}{\includegraphics{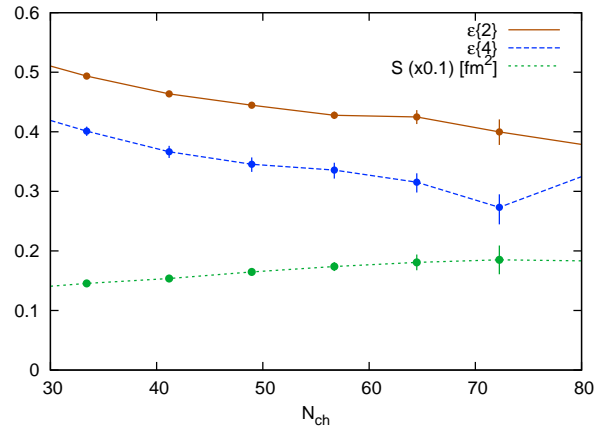}}}
\caption{(Color online)
Predictions for the eccentricity and the interaction area $S \, [{\rm fm}^2]\,(\times 0.1)$ at 7 TeV versus the  charged multiplicity in the interval $|\eta|<0.9$ at $\sqrt{s}=7$~TeV.}
\label{fig:epsilon}
\end{figure}

 In Fig.~\ref{fig:epsilon}, we plot the results for $\epsilon\{2\}$, $\epsilon\{4\}$ and $\langle S\rangle$  as a function of  $N_{ch}$ within the ALICE acceptance
$|\eta|<0.9$ (central detector) at 7 TeV. Events with $N_{ch}=60$ typically have 12 dipole--dipole (gluon--gluon) subcollisions. We see that the eccentricity is 20--40\% in the highest multiplicity region, similar to the value in semi--central nucleus--nucleus collisions.

\begin{figure}[h]
\resizebox{\linewidth}{!}{\rotatebox{-90}{\includegraphics{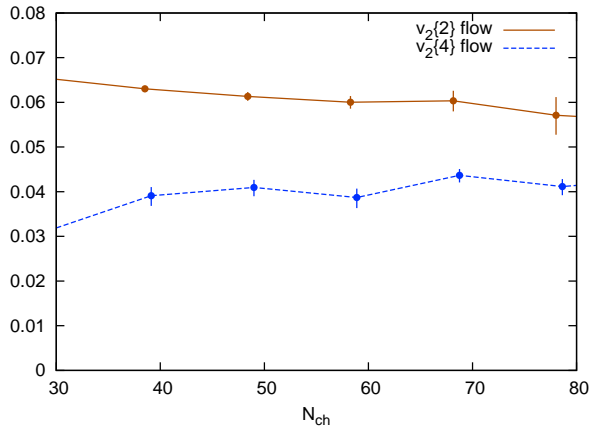}}}
\caption{(Color online) Predictions for $v_2\{2\}$ and $v_2\{4\}$ calculated from Eqs.~(\ref{hydroform2}) and (\ref{hydroform3})
versus the charged multiplicity in the interval $|\eta|<0.9$ at $\sqrt{s}=7$~TeV. The error bars are from statistics only. }
\label{fig:v2LHC}
\end{figure}

If the rescatterings among liberated partons (not included in DIPSY) are strong enough, then the
nonzero $\epsilon$ will give rise to $v_2$ according to the modified formulae (\ref{hydroform2}), (\ref{hydroform3}).
We plot the results for $v_2\{2\}$ and $v_2\{4\}$ in Fig.~\ref{fig:v2LHC}.
We see that $v_2$ is more or less constant and is about 4--6$\%$. This is comparable to the value in nucleus--nucleus collisions  at RHIC and at the LHC.
  We discuss the implications of these results below.
 The Monte Carlo shows no significant difference at 14 TeV (results not shown) even though high--multiplicity events
are then more frequent.

It is worth noting  that DIPSY predicts that the conventional definition of the eccentricity
\beq
\epsilon_s \equiv \frac{\sigma_y^2-\sigma_x^2}{\sigma_y^2+\sigma_x^2}\,,
\eeq
typically
takes a negative value, if the impact parameter is chosen in the
$x$--direction. This is in stark contrast to the nucleus--nucleus case where the interaction region is roughly the geometrical overlap of two colliding nuclei so that $\epsilon_s >0$ always. While $\epsilon_s$ is
unmeasurable in $pp$ collisions, this still illustrates the fact that the origin of the eccentricity is very different from that in the nucleus--nucleus case.


\section{Discussion}
\label{s:nonflow}


The rather large value of elliptic flow $v_2\sim 6\%$ that we have obtained in the flow scenario may seem promising at first. However, in practice the observed $v_2\{2\}$ and $v_2\{4\}$ differ from the genuine $v_2$ by the so--called nonflow contribution
\beq
(v_2\{n\})^n = v_2^n + \delta_n\,,
\eeq
 where $\delta_n$ is the $n$--particle correlations not associated with flow, such as resonance decays and the back--to--back correlations from hard and semi-hard scatterings. In nucleus--nucleus collisions, they are relatively innocuous because they scale with the multiplicity as
 \beq
 \delta_n \sim \frac{1}{N_{ch}^{n-1}}\,. \label{scale}
 \eeq
In $pp$ collisions, we expect that the nonflow contribution is less suppressed than (\ref{scale})
due to various initial and final state effects. Most importantly, high--multiplicity events often contain several jets and mini-jets, and particles within a pair of recoiling jets typically give a large contribution $\cos 2(\phi_i-\phi_j) \approx 1$ to the average (\ref{v2def}).
There are also correlations from the initial state partonic evolution which is initiated by only a few partons.
These correlations tend to enhance the nonflow contribution $\delta_n$, making the isolation of
$v_2^n$ difficult for small values of $n$.

Indeed, the ALICE collaboration has found that $v_2\{2,4\}$ decrease slowly with $N_{ch}$ \cite{Bilandzic:2011zz}. This slow decrease is also observed in Monte Carlo simulations, meaning that the scaling (\ref{scale}) does not hold
for $pp$ collisions. In the highest multiplicity events, $v_2\{2\} \approx 0.13$ \cite{Bilandzic:2011zz}  which is twice as large as the flow contribution $v_2\approx 0.06$. This implies that the two--particle correlation is dominated by nonflow effects.

This situation in $pp$ collisions necessitates us to look at higher order cumulants $v_2\{n\}$ with $n\ge 4$ which are by definition insensitive to  two--particle nonflow correlations.
 In this regard it is very interesting to notice that the ALICE collaboration \cite{Bilandzic:2011zz} also reported the measurement of $v_2\{4\}$ in $pp$, and found that the right--hand--side of (\ref{v4def}) is {\it negative}. [$(v_2\{4\})^4 \approx -10^{-4}$ in the highest multiplicity bins.] The same phenomenon can be seen in MC simulations like  PYTHIA (as done in \cite{Bilandzic:2011zz}) and also in DIPSY without assuming flow. On the other hand, as shown in Figs.~\ref{fig:epsilon} and \ref{fig:v2LHC}, $(\epsilon\{4\})^4$ and  $(v_2\{4\})^4$ are positive in the flow scenario at large $N_{ch}$. Leaving the origin of negative four--particle nonflow  correlations for future work, we conclude with a prediction that, if there is flow in the large $N_{ch}$ region, then the fourth order cumulant \eqref{v4def},
which is negative in the absence of flow, will  eventually turn positive.\\

\section*{Acknowledgements}
We are grateful to G\"osta Gustafson and Leif L\"onnblad for allowing us to use DIPSY, prior to publication \cite{Flensburg:2011kk}, for this particular observable.
We thank Alejandro Alonso,  Raimond Snellings, Julia Velkovska and the participants of  the workshop
``High Energy Strong Interactions 2010''  at Yukawa Institute in Kyoto University, for discussions.
 The work of Y.~H. and T.~U. is supported by Special Coordination Funds for Promoting Science and Technology of the Ministry of Education, the Japanese Government. E.~A. is supported by the US D.O.E. grant number
DE-FG02-90-ER-40577.


\begin{thebibliography}{99}


\bibitem{Ackermann:2000tr}
  K.~H.~Ackermann {\it et al.}  [STAR Collaboration],
  Phys.\ Rev.\ Lett.\  {\bf 86}, 402 (2001)
  [arXiv:nucl-ex/0009011].

\bibitem{Adler:2003kt}
  S.~S.~Adler {\it et al.}  [PHENIX Collaboration],
  Phys.\ Rev.\ Lett.\  {\bf 91}, 182301 (2003)
  [arXiv:nucl-ex/0305013].

\bibitem{Alver:2006wh}
  B.~Alver {\it et al.}  [PHOBOS Collaboration],
  Phys.\ Rev.\ Lett.\  {\bf 98}, 242302 (2007)
  [arXiv:nucl-ex/0610037].

\bibitem{Ollitrault:1992bk}
  J.~Y.~Ollitrault,
  Phys.\ Rev.\  D {\bf 46}, 229 (1992).





\bibitem{Drescher:2007cd}
  H.~J.~Drescher, A.~Dumitru, C.~Gombeaud and J.~Y.~Ollitrault,
  Phys.\ Rev.\  C {\bf 76}, 024905 (2007)
  [arXiv:0704.3553 [nucl-th]].

\bibitem{Luzum:2009sb}
  M.~Luzum and P.~Romatschke,
  Phys.\ Rev.\ Lett.\  {\bf 103}, 262302 (2009)
  [arXiv:0901.4588 [nucl-th]].

\bibitem{Bautista:2009my}
  I.~Bautista, L.~Cunqueiro, J.~D.~de Deus and C.~Pajares,
  J.\ Phys.\ G {\bf 37}, 015103 (2010)
  [arXiv:0905.3058 [hep-ph]].

\bibitem{Chajecki:2009es}
  Z.~Chajecki and M.~Lisa,
  Nucl.\ Phys.\  A {\bf 830}, 199C (2009)
  [arXiv:0907.3870 [nucl-th]].

\bibitem{Prasad:2009bx}
  S.~K.~Prasad, V.~Roy, S.~Chattopadhyay and A.~K.~Chaudhuri,
  Phys.\ Rev.\  C {\bf 82}, 024909 (2010)
  [arXiv:0910.4844 [nucl-th]].

\bibitem{d'Enterria:2010hd}
  D.~d'Enterria, \emph{et al., }
  Eur.\ Phys.\ J.\  C {\bf 66}, 173 (2010)
  [arXiv:0910.3029 [hep-ph]].

\bibitem{Bozek:2009dt}
  P.~Bozek,
  Acta Phys.\ Polon.\  B {\bf 41}, 837 (2010)
  [arXiv:0911.2392 [nucl-th]].

\bibitem{Ortona:2009yc}
  G.~Ortona, G.~S.~Denicol, Ph.~Mota and T.~Kodama,
  arXiv:0911.5158 [hep-ph].




\bibitem{CasalderreySolana:2009uk}
  J.~Casalderrey-Solana and U.~A.~Wiedemann,
  Phys.\ Rev.\ Lett.\  {\bf 104}, 102301 (2010)
  [arXiv:0911.4400 [hep-ph]].

\bibitem{Pierog:2010wa}
  T.~Pierog, S.~Porteboeuf, I.~Karpenko and K.~Werner,
  arXiv:1005.4526 [hep-ph].

\bibitem{Aamodt:2010pp}
  K.~Aamodt {\it et al.}  [ALICE Collaboration],
  Eur.\ Phys.\ J.\  C {\bf 68}, 345 (2010)
  [arXiv:1004.3514 [hep-ex]].

\bibitem{Collaboration:2010gv}
  CMS~Collaboration,
  JHEP09(2010)091
  [arXiv:1009.4122 [hep-ex]].

\bibitem{Dinh:1999mn}
  P.~M.~Dinh, N.~Borghini and J.~Y.~Ollitrault,
  Phys.\ Lett.\  B {\bf 477}, 51 (2000)
  [arXiv:nucl-th/9912013].



\bibitem{Ollitrault:2009ie}
  J.~Y.~Ollitrault, A.~M.~Poskanzer and S.~A.~Voloshin,
  Phys.\ Rev.\  C {\bf 80}, 014904 (2009)
  [arXiv:0904.2315 [nucl-ex]].

\bibitem{Avsar:2005iz}
  E.~Avsar, G.~Gustafson and L.~L\"onnblad,
  JHEP {\bf 0507}, 062 (2005)
  [arXiv:hep-ph/0503181].

\bibitem{Avsar:2006jy}
  E.~Avsar, G.~Gustafson and L.~L\"onnblad,
  JHEP {\bf 0701}, 012 (2007)
  [arXiv:hep-ph/0610157].

\bibitem{Avsar:2007xg}
  E.~Avsar, G.~Gustafson and L.~L\"onnblad,
  JHEP {\bf 0712}, 012 (2007)
  [arXiv:0709.1368 [hep-ph]].


\bibitem{Flensburg:2008ag}
  C.~Flensburg, G.~Gustafson and L.~L\"onnblad,
  Eur.\ Phys.\ J.\  C {\bf 60}, 233 (2009)
  [arXiv:0807.0325 [hep-ph]].

\bibitem{Flensburg:2011kk}
  C.~Flensburg, G.~Gustafson and L.~Lonnblad,
  arXiv:1103.4321 [hep-ph].

\bibitem{Flensburg:2010jk}
  C.~Flensburg,
  arXiv:1009.5323 [hep-ph].







  \bibitem{Mueller:1993rr}
  A.~H.~Mueller,
  Nucl.\ Phys.\  B {\bf 415}, 373 (1994).

\bibitem{Mueller:1994gb}
  A.~H.~Mueller,
  Nucl.\ Phys.\  B {\bf 437}, 107 (1995)
  [arXiv:hep-ph/9408245].

\bibitem{Salam:1995zd}
  G.~P.~Salam,
  Nucl.\ Phys.\  B {\bf 449}, 589 (1995)
  [arXiv:hep-ph/9504284].

\bibitem{Hatta:2007fg}
  Y.~Hatta and A.~H.~Mueller,
  Nucl.\ Phys.\  A {\bf 789}, 285 (2007)
  [arXiv:hep-ph/0702023].

\bibitem{Avsar:2008ph}
  E.~Avsar and Y.~Hatta,
  JHEP {\bf 0809}, 102 (2008)
  [arXiv:0805.0710 [hep-ph]].




\bibitem{Lonnblad:1992tz}
  L.~L{\"o}nnblad,
  Comput. Phys. Commun.{} {\bf 71},~15~(1992).

\bibitem{Sjostrand:2006za}
  T.~Sjostrand, S.~Mrenna and P.~Z.~Skands,
  JHEP {\bf 0605}, 026 (2006)
  [arXiv:hep-ph/0603175].



\bibitem{Andersson:1983ia}
  B.~Andersson, G.~Gustafson, G.~Ingelman, and T.~Sjostrand,
  Phys. Rept.{} {\bf 97},~31~(1983).


\bibitem{Bhalerao:2006tp}
  R.~S.~Bhalerao and J.~Y.~Ollitrault,
  Phys.\ Lett.\  B {\bf 641}, 260 (2006)
  [arXiv:nucl-th/0607009].

\bibitem{Qiu:2011iv}
  Z.~Qiu and U.~W.~Heinz,
  arXiv:1104.0650 [nucl-th].


\bibitem{Bilandzic:2011zz}
  A.~Bilandzic  [ALICE Collaboration],
  AIP Conf.\ Proc.\  {\bf 1343}, 465 (2011);
A.~Bilandzic,  talk given at
`Quark Confinement and the Hadron Spectrum IX',
\url{http://147.96.27.42//contributionDisplay.py?contribId=110&sessionId=124&confId=0}


\end{thebibliography}
\end{document}